\documentclass[
    ,final            
  ]
  {aipproc}

\layoutstyle{8x11single}
\usepackage{epsfig,graphicx,times,amsmath,amsthm, amssymb, multicol}
\usepackage[T1]{fontenc}  
\usepackage{calc}
\usepackage{amsmath}
\usepackage{amsfonts}
\usepackage{amssymb}
\usepackage{multirow}

\newcommand{\mb}{\mathbf }
\newcommand{\mc}{\mathcal}
\newcommand{\pr}{^\prime}
\newcommand{\usd}{\textit{u.s.d.}}

\begin{document}

\title{Fourier Heat Conduction as a phenomenon described within the scope of the
Second Law}


\author{Christopher G. Jesudason}{
  address={Chemistry Department and Center for Theoretical and Computational
Physics\\
	University of Malaya, 50603 Kuala Lumpur, Malaysia} }
\begin{abstract}
The historical development of the Carnot cycle  necessitated  the construction
of   isothermal and adiabatic pathways within the cycle that were also 
mechanically "reversible"  which lead eventually to the  Kelvin-Clausius
development of the entropy function $S $ where for any reversible closed path
$\mc{C}$, $\oint_{\mc{C}} dS =0$ based on an  infinite number of concatenated
Carnot engines that approximated the said path and  where for each engine
$\Delta Q_1/T_1 + \Delta Q_2/T_2 = 0$ where the Q's  and T's are the heat
absorption increments and temperature respectively with the subscripts
indicating the isothermal paths $(1,2)$ where for the Carnot engine, the heat
absorption is for the diathermal (isothermal)  paths  of the cycle only. Since
'heat'   has been defined  as that form of energy that is transferred as  a
result of a temperature difference  and a corollary  of the Clausius statement
of the Second law is that it is impossible for heat to be transferred from  a
hot to a cold reservoir with no 
other effect on the environment, these statements suggested  that the local mode
of transfer of 'heat' in the isothermal segments of the pathway  does imply a
Fourier heat conduction mechanism (to conform to the definition of 'heat')
albeit of a ``reversible'' kind,  but on the other hand, the Fourier mechanism
is  apparently irreversible, leading to an increase in entropy of the combined
reservoirs at  either end of the material  involved in the conveyance of  the
heat energy. These  and several other considerations lead Benofy and Quay (BQ) 
to postulate the  Fourier heat conduction phenomenon to be an ancillary
principle in thermodynamics, with this principle being strictly local in nature,
where the global Second law statements could not be applied to this local
process. Here we present equations that model heat conduction as a
thermodynamically reversible but mechanically irreversible process where due to
the belief in mechanical time reversible symmetry, thermodynamical reversibility
has been 
unfortunately linked to mechanical reversibility, that has discouraged such an
association. The modeling is  based on an application of a ``recoverable 
transition'', defined and developed earlier on ideas derived from thermal
desorption  of particles from a surface  where the Fourier heat conduction
process  is approximated as a series of such desorption processes.  We recall
that the original Carnot engine required both adiabatic and isothermal steps to
complete the zero entropy cycle, and this construct lead to the consequent
deduction that  any Second law statement that refers to heat-work conversion
processes are only globally relevant. Here, on  the other hand,  we examine 
Fourier heat conduction from MD simulation and model this process as a
zero-entropy forward scattering process relative to each of the atoms in the
lattice chain being treated as a system where the Carnot cycle can be applied
individually. The equations developed predicts the  ``work'' done  to be equal
to the energy transfer rate. 
The MD simulations  conducted  shows excellent   agreement with the theory. Such
views  and results as these, if developed to a successful conclusion could imply
that  the Carnot cycle be viewed as describing a local process of energy-work
conversion and that   irreversible local processes might be brought within the
scope of this cycle, implying a unified treatment of  thermodynamically (i)
irreversible, (ii) reversible, (iii) isothermal and (iv) adiabatic processes.
\end{abstract}

\maketitle


\section{Introduction}\label{sec:1}
Some principles were developed   in connection with the  thermal desorption
problem, where it was proven  that the irreversible desorption of particles on a
surface \cite{cgj14} within a relaxation time $\tau$ conforms to an adiabatic
scattering zero entropy process along the center-of-mass frame of the desorbing
particles. To summarize the more detailed treatment in \cite[Sec. 4.4]{cgj14},
for states $i$ with probability $p_i$ before a transition and $p^\prime_i$ after
where state i transmutes  into another state (denoted by primes), the entropy
change $\Delta S$
\begin{equation}\label{eq:1}
	\Delta S=kk^\prime\Delta \sum_{i=1}^N{p_i\ln p_i}=0 
\end{equation}
 for the entire transition implied 
\begin{eqnarray}\label{eq:2}
	p_i(E_i,T)&=&p\pr_i(E\pr_i,T\pr) \\
	E_i/(kT)&=&E\pr_i/(kT\pr)
\end{eqnarray}
where the $E$'s are energy variables   between the two  states with defined
temperatures $T,T\pr$ and where the partition function  $\mc{Z}$ too is
conserved, i.e. $\mc{Z}=\mc{Z}\pr$ , and where if we summed the energies in
(\ref{eq:2}), we have the heat energy $Q$ conforming to 
\begin{equation}\label{eq:3}
	\frac{Q}{kT}- \frac{Q\pr}{kT\pr} = 0 
\end{equation}
or the entropy is conserved  for this ``heat pulse'' movement called a
recoverable trajectory. We note that the prime and unprimed variables refer to
the same system making a  transition. These transitions are ``thermodynamically
reversible'' in the sense of zero entropy change, but is clearly mechanically
irreversible \cite{cgj16}.
For the above system, we  observe that the system is disintegrating, and that
the ad-atoms on the surface of the substrate eventually all leave the surface,
and each of these ad-atoms have instantaneous  energy $E\pr$,where  $E\pr = E -
\delta \mc{W}$ where $\delta \mc{W}$ is the work done by the particle on the
force-field it traverses, which is absorbed into the field. The particles then
move on to infinity in space if no work potentials are present  along the
trajectory when the kinetic energy of the particles  are  still positive. 
Clearly in this situation, no structure is preserved. Of interest then is to 
investigate the possibility of modeling energy transfer processes that has
something of the form above of zero entropy trajectories relative to  the
defined heat terms, but where the system is not disintegrating. A simple example
of such a structured assembly would be a lattice chain where the particles have
a mean time independent position whilst involved in the process of transferring
energy 
between the 
two ends of the lattice chain that are maintained at different temperatures.
Since  modeling such a process is a very involved task, with possibly unproven
assumptions being  surreptitiously imported into the creation of the theoretical
structure that  could be simulated and computed,  the following methodology is
being currently  pursued to ensure that no extraneous concepts are imported:
 \begin{enumerate}
	 \item Identify a  discrete non-disintegrating model system  that, from
theoretical considerations would have to conform to the above process of a
recoverable process \label{en:1}
	\item Apply that model to the continuous lattice chain, even in a
restricted system to extend concepts that would have to be used for a more
comprehensive treatment of irreversible phenomena \label{en:2}
	\item quantitatively check the model for numerical agreement with total
heat flow terms \label{en:3}
	\item  extend the model from just one particle interaction  with the
adjacent pair to longer ranges of interactions so that one might be able to
predict the kinetics and thermodynamical variable profile over the entire
lattice \label{en:4}
	\item Lastly, from the above, it may be possible to construct a more
comprehensive and extensive  thermodynamical theory by generalization of the
above item (\ref{en:4})    that might be able to encompass both  equilibrium and
non-equilibrium interactions. \label{en:5}
 \end{enumerate}
So  far we  seem to have  promising results  for   items (\ref{en:1}-\ref{en:3})
which will be discussed here. The other items in the methodology list 
(\ref{en:4}-\ref{en:5}) are work in progress.  

 It was demonstrated before \cite{cgj16,cgj4,cgj5} that the pivotal concept of
time reversibility is   mathematical incorrect in its major applications  to 
wide-ranging phenomena, and that these fallacies have been incorporated into
mainstream thermodynamical interpretation (e.g. see
\cite[pp.36,52,141-7,356-7]{lands1} ). The above  remarks require some
qualification. The original work of Carnot \cite{carnot1} describes essentially
a work cycle  whose work energy variable $w$  is not a perfect differential and 
where the introduction of heat \cite[Fig. 2,3 p.70]{carnot1} into the working
substance involves the transfer of heat about vanishing temperature gradients
where there is  ``... no contact between bodies of sensibly different
temperatures''\cite[p.68]{carnot1}. Such views developed in modern times to
conceiving heat as a degraded form of energy that increased the entropy of a
system by traversing a thermal gradient from hot to cold (thus increasing the
entropy of the system). Indeed the 
calorimetric definition of heat \cite[1968 5th Ed, p.73]{zemansky1} is ``...that
which is transferred between a system and its surroundings by virtue of a
temperature difference only.''  Carath\'{e}odory conceives of heat  \cite[J.
Kestin ed., Introduction, p.229]{cara1} in the following manner :
{\em"`Furthermore, when two bodies of different temperatures are brought into
contact, heat always passes from the hotter to the colder, and never in the
reverse direction."'}. Here we shall show for  recoverable transitions, which
includes heat conduction, the energy  transferred is actually ``work'', although
the movement is from hot to cold, in accordance with previous definitions.  The
degradation idea is also evident in  \'{E}. Clapeyron's 1834 work 
\cite[p.38]{clap1}:''...in any mechanism designed to produce motive power from
heat,there is a loss of force   whenever  there is a direct communication of
heat between two bodies at different temperatures and it follows that the
maximum effect can be produced only 
by a mechanism in which contact is made only  between bodies at equal
temperatures''.  This coupled with  the definition of heat above lead to the
concept of  ``reversible transfer'', where the concept of ``reversibility'' is a
rationalization imported from mechanics with its belief in the reversibility of
its laws in quantum and classical physics and thermodynamics \cite{lands1}. The
derivation of the Onsager reciprocity relations and the Boltzmann H-theorem  are
examples  of  misapplications of such time-reversible assumptions
\cite{cgj16,cgj4,cgj5}.

Finally, it was proven that the celebrated Liouville equation, derived from the
Hamiltonian in $(\mb{p},\mb{q})$  Liouville phase space is  in general not valid
as a continuous equation, where a stochastic analog  of the same form was
derived in its place \cite{cgj12}. In particular, attempts  to deduce
zero-entropy paths from Liouville space were shown to be flawed \cite[see
reference therein]{cgj12}. It was expressly stated  in \cite{cgj14} that the
zero-entropy ``recoverable trajectory'' developed there  is \textit{not}
described in Liouville space, and that further, as illustrated in  the current
work, there is  a stochastic back-scattering of energy involving a
non-conservation  of energy about a stochastic work cycle that at first sight
would not readily follow from a  standard mechanical Hamiltonian using
continuous, non-stochastic variables as will be demonstrated. If such
Hamiltonians $\mc{H}(\mb{p,q})$ are  used in describing non-equilibrium
mechanical systems, then assuming a  general average $\{\
mb{\bar{p}} ,\mb{\bar{q}}\}$
for all coordinates $j$ where $j \ne i$, then $\oint {\frac{\partial
\mc{H}}{\partial q_i}(q_i, \mb{\bar{p},\bar{q})} dq_i} =0$ which is  not
observed in the simulation  result,  implying that the introduction of hybrid
elements (random energy impulses at the ends of a chain of vibrating atoms in
this case) to for instance simulate  thermostated regions destroys the continuum
description of the mechanical Hamiltonian and that  in addition some very
complex cooperative phenomena may be  involved  with the standard Hamiltonian
where the assumption of averaged coordinates do not apply. The conventional 
description  of conductive heat  transfer  which is viewed as the transport of 
"heat"  energy  occurs through reversible dynamical laws, and  which involves a
positive local internal entropy production rate ($\int_V\sigma dV=dS_i/dt \geq
0$)\cite{degroot1}. If the entropy vector has a component  $\mb{J_s}=\mb{J_q}/T$
($\mb{J_q}$ being the heat current vector), then at the steady state
 \begin{equation}
\nabla .\mb{J_s}=-\mb{J_q}.\nabla T/T^2\neq 0  
	\label{eq:4}
	\end{equation}
	in a temperature gradient with a non-zero heat current present and
indeed in conventional descriptions, $\sigma$ has the above component due to the
conductive heat contribution \cite[Ch.III,p.24, eq. (21)]{degroot1}. One
important aspect  of the expressions for $\sigma$  is that it allows for the
identification of the Onsager reciprocity coefficients that couple forces and
fluxes \cite[Ch.IV,pp.33-36]{degroot1}.  Previously, it was remarked
\cite[p.163,Sec.2]{cgj14}, in keeping with the Clausius definition, that for
closed systems a more appropriate definition would be $\dot{S}=-\int_V
\nabla.\mb{J_q}/T$ which would yield zero entropy production apart from the
system boundaries. Further, Benofy and Quay have argued that Fourier's
inequality (following (\ref{eq:4}) with $\mb{J_q}=-\kappa\nabla T$)
\begin{equation}
\kappa(\nabla T)^2/T^2 \geq 0
\label{eq:5}
\end{equation}
is a local principle, not subject to the Second law since the latter refers to
global work-heat transitions, and may well be an independent principle. Recall
that the modern Carnot cycle is strictly developed with 2 types of heat
transitions of the working substance, the "`isothermal"' and the "`adiabatic"'.
One conclusion of Carath\'{e}odory thermodynamics \cite{cara2} for
"`macroscopic"' systems is found in Axiom II:\,{\em In every arbitrary close
neighborhood of a given initial state, there exists states that cannot be
approached arbitrarily closely by adiabatic processes.} In what follows, we show
that relative to a particular  subsystem, within the  heat conducting chain 
constituting  atom $i$ and the adjacent one $i+1$, pure heat energy transfer
according to Carath\'{e}odory occurs  and in accord with  conventional
definition, but at the same time according to the major assumptions of
recoverable transitions \cite{cgj14} and as illustrated here in our modeling,
the following conditions obtain:

\begin{enumerate}
\item
  A net adiabatic process in the forward direction occurs
\item
  The energy transfer being the work done, conventionally  considered the heat
increment
  
  \item
   The work is mechanically irreversible,  but thermodynamically reversible
(zero entropy)  where the micro-transitions complete a Carnot-type loop within
the subsystem
  \item
   The stochastic integral for the energy about a loop is not zero, implying
that the mechanical Hamiltonian is non-conservative in such hybrid systems
\item
  There is a distant implication that both adiabatic and isothermal processes
may be unified, at least at the micro-level, which would require answering
challenging questions as to how macroscopic descriptions, such as due to
Carath\'{e}odory , emerge from the micro-processes, which encompasses both
adiabatic and isothermal ones.
\end{enumerate}

The above seems to indicate that  if the direction of this research project is
deemed reasonable, then there are gaps that need to be bridged between some of
the more conventional definitions and the deductions that are being made here.

\section{Description of  the simulation system}\label{sec:2}
The sample results presented here refers to a 1000 atom chain, labeled 1 to 1000
from left to right, with the first 200 atoms  on the left thermostated to 4.0
(reduced units)  whilst atoms 800-1000 were maintained at 1.0 . The method of
thermostating used a classical, non-synthetic algorithm developed or popularized
by Hafskjold and  Ikeshoji \cite{Haf7}where the  thermostated atoms were scaled
according to $\dot{q}\pr_i=\mb{\alpha}+\mb{\beta}\dot{q}_i$, with $\mb{\alpha}$ 
and $\mb{\beta}$ common to all relevant  atoms to maintain the temperature $T$
where $ T=\sum_{i=N_a}^{N_b}{m\dot{q}^2}$ and in reduced units, $m=1$. As is
well know, the  harmonic potential only  \cite{rieder1} does not yield the
expected Fourier heat conduction law $\mb{J_q}=-\kappa \nabla T$, with near
constant $\kappa$, but one might also say that for the harmonic interaction
potential between particles, the thermal conductivity  $\kappa$ is a   very
sensitive  function of the temperature, and that it is also not a local property
but 
may be  a function of the entire temperature distribution. These  and others are
very interesting  research  questions, as Lebowitz et al.  have  testified to
\cite{lebo2}. The interparticle potential $\mc{V}$ between particles $i$ and
$i+1$  was defined as 
\begin{equation}\label{eq:6}
	\mc{V} = k_h(q_{i+1}- q_{i})^2/2 +b_h(q_{i+1}- q_{i})^4/4 .
\end{equation}

Here $k_h=1.0,b_h=0.5$ with these parameter values chosen for good
reproducibility in the work on Tejal et al. \cite[Fig.6]{tejal1}. The $q$'s are
the displacement from the equilibrium position with the separation distance of
unity, and the force on particle  $i$ due to particle $i+1$ is defined as
$F_{i,i+1}=-\frac{\partial \mc{V}}{\partial q_i}$. We define the {\em
partitioned work} done \textit{on}  $i$ due to the force from $i+1$ as 
\begin{equation} \label{eq:7}
	\Delta w_{i+1 \rightarrow i} = \int_{t_1}^{t_2} F_{i,i+1}\, \dot{q}_i
\,dt \approx m\oint_{stoch} F_{i,i+1} \,dq_i
\end{equation}
between the time interval $[t_1,t_2]$, and the work done \textit{on} $i+1$ due
to the force from $i$ as 
\begin{equation} \label{eq:8}
	\Delta w_{i \rightarrow i+1} = \int_{t_1}^{t_2} F_{i+1,i}\,
\dot{q}_{i+1} \,dt \approx m\oint_{stoch} F_{i+1,i} \,dq_{i+1}.
\end{equation}
 The above equations normalized over unit time are defined thus:
\begin{eqnarray}\label{eq:8b}
	\delta w_{2\rightarrow 1 }=\Delta w_{i+1 \rightarrow i}/(t_2-t_1)  \\
	  \delta w_{1\rightarrow 2 }=\Delta w_{i \rightarrow i+1}/(t_2-t_1) . 
\label{eq:7b}
\end{eqnarray}
We make use of the standard Simpson second order 3-point formula (fourth order 
error) numerical integration \cite{nrc,yak1} for the  over 3 million (M)
consecutive  time steps of stepsize  $0.001$, equal to the MD timestep. Since
all the particles are oscillating, particles $i$ and $i+1$ too can be viewed as
oscillating back and forth an average of  approximately $m$ times (to the
nearest integer) about their mean position. If this partition force were
conservative, then 

\begin{eqnarray}\label{eq:9}
	 \lim_{m \to \infty } \Delta w_{i+1 \rightarrow i}/m&=&0 \\
	  \lim_{m \to \infty } \Delta w_{i \rightarrow i+1}/m &=&0
\end{eqnarray}
and the above limits for  large values of time intervals are   not observed  in
all the simulations carried out (see Table (\ref{tab:1}) )  with varying 
lengths from 3 to 10 M time step intervals, where, when these integrations of
different time lengths  are  normalized to per unit time, yielded  the same
numerical quantities.  Hence these stochastic path integrals (involving a hybrid
system of random energy and momentum impulses at the reservoir coupled to a
standard system Hamiltonian without a time  dependent variable)      have
non-conservative "`Hamiltonians"' even if the classical Hamiltionian is 
continuous  with  continuous variables having  no explicit time dependence. We
note that  distinguished theorists routinely use the Liouville  and Hamilton 
equations in modeling  these heat conduction problems  \cite[etc.]{rieder1,
lebo2} etc. .

\section{Formulation of theoretical reference model}\label{sec:3}
Assuming the validity of equipartition, this particular model  will  exhibit
recoverable transitions  in the manner described below. Of note here is that the
localized particle exhibits two temperatures as it interacts with adjacent
particles; there is the mean kinetic energy temperature and the temperature
associated with the transfer  of energy  to the adjacent particle (on the right,
particle $i+1$ ) in a basic adiabatic transition. Once the model is described,
then we actually use it  for the continuously interacting  case that we validate
by simulation; the key link to the continuum is to realize that  each moment of
time constitutes a ``collision'' interaction of the coupled particles relative
to this "`isolated"' reference model. 

\begin{figure}[ht]\label{fig:2} 

	\centering
width=6cm]{tp1.eps}
	\scalebox{0.5}{\includegraphics[70mm, 170mm] [160mm, 250mm] {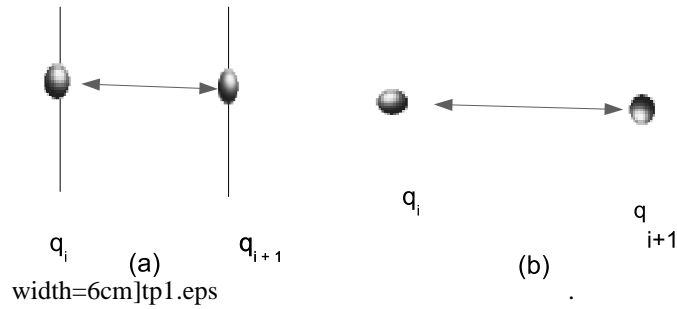}}
  \caption{Schematic  (a) represents  adjacent particles that interact by hard
sphere elastic collisions where a harmonic potential operates about the
equilibrium position with no direct  harmonic or other force coupling between
these particles  except for the hard sphere collisions; (b) represents  a system
where the adjacent  particles interact continuously and directly though an
interparticle  potential function, such as given in eq.(\ref{eq:6})}.
\end{figure}

The theoretical  model  is ``isolated'' in the sense that its energy is well
defined and localized except for the time of collision with elastic hard sphere
energy interchange  as shown in Figure \ref{fig:2}(a) where  they oscillate in
the horizontal direction due to a potential $V$ acting  vertically with a small
horizontal  projection which we can assume to be harmonic with regard to its
displacement $q_i$ from the horizontal equilibrium position at $q_i=0$, (this
assumption is not   mandatory but it simplifies matters)  and so $V_i=\frac{k_h
q_i^2 }{2} $, the kinetic energy is $k.e.(i)=\frac{m \dot{q}_i^2 }{2} $ and the
total energy $E_i=V_i + k.e.(i)$. We will eventually modify this model to cover
the situation in Figure \ref{fig:2}(b) by appropriate choice of subsystem where 
even in the coupled state, one can conceptualize each particle as being isolated
at each instance of time. For what follows, we  refer to particle $i$  simply as
$i$. The proof of the workability  of this model is that a simulation 
of an allied system with the same dynamics has been described in detail
\cite{prosen1}.
The temperature $T_i$ of  $i$ from equipartition may be defined as an ensemble
average 
\begin{equation}\label{eq:10}
	kT_i=\left\langle k.e.(i) + V_i\right\rangle = \left\langle
E_i\right\rangle .
\end{equation}
The above (\ref{eq:10}) is computable. The equilibrium distance between
particles  is set to $1$ in reduced units. Let $pl_j$  denote both the plane
perpendicular to the vector $q_i-q_{i+1}$ which contains the $q$ coordinate
point of contact between the particles during the $jth$ collision of $i$ and
$i+1$,  and the $q$ coordinate during this collision process. The collision
would impart a change of kinetic energy $\delta \dot{q}_i$. By energy
conservation, the average energy up to the time of collision for $i$,  $Q_b$ is 

\begin{equation}\label{eq:11}
	Q_b=\left\langle k.e.(i) + V_i|_{pl_j}\right\rangle (\mbox{averaged over
all j collisions}) .
\end{equation}

We note that $Q_b$ is also dependent on interaction with particle $i-1$ which is
not relevant here for the transfer  energetics of $i$ to $i+1$. Then the
dynamical laws can be utilized to compute the change of k.e. for $i$ (since the
potential energy remains unchanged at $q_i=pl_j$ and we can compute $Q_a$ the
energy just \textit{after} the collision as 
\begin{equation}\label{eq:12}
	Q_a=\left\langle k.e.\pr(i) + V_i\pr|_{pl_j}\right\rangle
(\mbox{averaged over all j collisions})
\end{equation}
over  the time  duration between the  previous  collision $j-1$ denoted $\Delta
t$; we define the mean time between collisions 
as $\tau_{mc}$.   Averaging over the sampling time, we can compute average
quantities such as  the  $Q$'s , denoted by brackets  $<>$ and so over 
time $\tau_{mc}$, the energy transfer $\delta w$ is 
\begin{equation}\label{eq:13}
	\delta w=<Q_b>-<Q_a>
\end{equation}
or $\delta w/\tau_{mc}$ per unit time over a continuous time period.  Note that
the normalized  units used here are not the same as 
for the un-normalized work terms in  (\ref{eq:7}-\ref{eq:8}).
From (\ref{eq:10}),  the following array of equations become evident:
\begin{eqnarray}
T_b &=& k<E_b>  =k<Q_b> \\
T_a &=& k<E_a>  =k<E_a -\delta w> \\
T_a &=& k<E_a>  =k<E_b + \delta E> = k<Q_a>\\
\frac{1}{k}&=& \frac{Q_a}{T_a}=\frac{Q_b}{T_b} .
\end{eqnarray}
From one of the field properties of number theory \cite[p.15, Field Axiom
A4]{depree1}, 
\begin{eqnarray}
 \frac{1}{k} + \left(-\frac{1}{k}  \right)&=& 0  \label{eq:14} \\
  &\mbox{we deduce }&   \nonumber     \\
\delta S= \frac{<Q_b>}{T_b} -\frac{<Q_a>}{T_a} &=& 0 \label{eq:15}
\end{eqnarray}
which defines the recoverable trajectory. Some elementary remarks are in order
concerning the averaging process. We note that 
\begin{equation}
<Q_a>=\left(\sum_{i=1}^N Q_{a,i}/N\right)
\label{eq:beta}
\end{equation}
Then from ( (\ref{eq:14}) or (\ref{eq:15})  ) and some $k\pr$ (here $k=k\pr$),
\begin{eqnarray}
<T_a>=\frac{k\pr\sum Q_{a,i}}{N}&=&k\pr<Q_a> \\
\Rightarrow \,\,\frac{<Q_b>}{<T_b>} - \frac{<Q_a>}{<T_a>} &=& 0 \label{eq:16}
\end{eqnarray}
and the mean rate of energy transfer is $\delta w/\tau_{mc}$.
\subsection{A generalization of the Zeroth law based on recoverable trajectory}
The equilibrium understanding of the Zeroth law is that if two bodies are in
diathermal contact, and there is no net exchange of energy, then they are at the
same temperature. Another  entropic approach  is to examine  the exchange of
energy across  the diathermal boundary between two systems 1 and 2 (these labels
determine the heat and temperature of the systems) and if there is no net
exchange of energy $<\delta Q_1>=-<\delta Q_1>=0$, then  for any instant of time
$\delta Q_1=-\delta Q_1=0$ for non-work energy exchange, and 
\begin{eqnarray}
\frac{\delta Q_1}{T_1} + \frac{\delta Q_2}{T_2} &=&\frac{\delta Q_1}{T_1}- 
\frac{\delta Q_1}{T_2}, \\
\delta S=\frac{\delta Q_1}{T_1}-  \frac{\delta Q_1}{T_2} =0 &\Rightarrow& \delta
Q_1(T_2-T_1)=0
\end{eqnarray}
or $T_2=T_1$ if $\left|\delta Q_1\right|\neq 0 $. Since in dynamic equilibrium
$\left|\delta Q_1\right|\neq 0 $ most of the time, we conclude $T_2=T_1$. If we
therefore define the Zeroth law in terms of Zero entropy transitions, then
recoverable trajectories  $\delta(Q/T)=0$ can conform to the Zeroth law even if
their temperatures differ across a boundary if the criterion refers to  zero
entropy changes rather than energy (in at least closed systems). From the above
model, we note that the isolated system $i$ can be viewed as having
"`adiabatic"'-like transitions in terms of   energy  transfer at the time of
collision, leading to a double temperature characterization for the particle;
thus we can say that the two temperatures  refer to the same particle in
equilibrium with itself.

We have theoretically demonstrated that subsystem (a) of Figure {\ref{fig:2}
must have the particle behaving as a recoverable trajectory based on the
generalized equipartition theorem. Hence if the above model is applied to a
system that conforms to case (b) of the same Figure, and where there is
numerical confirmation that the vector "`work term"' $\delta \dot{w}\equiv\delta
w/\tau_{mc}$  in unit time arising from recoverability theory is exactly  equal
to the energy transfer across the  thermostated ends of the chain  in unit time,
 then one can conclude  that a verification of conductive heat as a recoverable
process has been made. We demonstrate this to be the case for the  9
atoms/particles  situated uniformly over the conductive lattice, where there is 
close numerical agreement between the independently determined heat transfer
rate due to the thermostats and the numerical integrations of various defined
energy processes for all of the above labeled particles. 
\section{Application of the Theoretical Reference Model described above  to
lattice particles interacting continuously via potentials}\label{sec:4}
Here we apply the results of the previous section  for discontinuous energy
interactions to continuous heat conduction (Figure \ref{fig:2} (b)). This  can
be achieved  by isolating the system to the "`particle"' with $coordinates
(q_i,\dot{q}_i)$ and where with $k/2=k\pr$ we have $k\pr<T_i>=<k.e.(i)>$. For
what follows we determine the stochastic averages omitting the  angle brackets
$<>$  normally indicative of this in the equations about (\ref{eq:16}), but the 
context should  indicate if averages are implied or not. We define the partial
heat to be $Q=k.e.(i)$ for (particle) $i$, where $k\pr T(i)=Q=k.e.(i)$. We then
correlate the reference system above to the current model by realizing that each
moment of time  involves an interaction via a potential, which takes the place
of the transfer of energy due to elastic collisions and the conservation of
momentum in the reference system.
Thus we can envisage an interaction over time interval $dt$, and for the $j$th
interval, the work performed on $(i+1)$ because of the force from $i$ is 
\begin{equation}
\delta w_j= F_{i+1,i}.\frac{dq_{i+1}}{dt}dt_j .
\label{eq:17}
\end{equation}
Then $\frac{Q_b}{T_b}=k\pr$ and after the interaction time $dt_j$, we have the  
following results where subscripts $a$ and $b$ refer to \textit{after} and
\textit{before} the interaction over time time interval $dt$ at the $j$th
interval  respectively:
\begin{eqnarray}
Q_a&=&(k.e.(i) -\delta w_j   )\\
k\pr T_a &=& ( k.e.(i) -\delta w_j)=Q_a \\
\mbox{and}\,\,\,k\pr&=&  k\pr \Rightarrow \nonumber \\
\frac{Q_b}{T_b}&-& \frac{Q_a}{T_a} =\delta S=0  \label{eq:18}
\end{eqnarray} 

where  (\ref{eq:18}) defines a recoverable process. Then over $n$  time
intervals, where $\Delta\mc{T}=ndt$,

\begin{equation}
<Q_b>=\int k.e.(i) dt/\Delta\mc{T}.
\label{eq:19}
\end{equation}
Define $\Delta \mathfrak{t}=dt$ and $F=F_{i+1,i}$. For any one time interval
$\Delta \mathfrak{t}$, the loss of $k.e.(i)$ which would  give it some of the
characteristics of $i+1$, such as a lower temperature  would also yield a heat
content $Q_a$  given as
\begin{eqnarray}
Q_a &=& k.e.(i) - F \frac{dq_{i+1} }{dt}dt_j  \label{eq:20}\\
\delta w &=& ( Q_b- Q_a) \,\,\,\,\mbox{spanning}\,\, [t_1,t_2].  \label{eq:21}
\end{eqnarray}}.

Over $n$ time intervals, the averaged $n \delta w$ has value
\begin{eqnarray}
n \delta w &=& \int_{t_1}^{t_2}  k.e.(i) dt/\Delta \mathfrak{t} -
\int_{t_1}^{t_2} (k.e.(i) -F 
\frac{dq_{i+1} }{dt} \Delta \mathfrak{t} )dt /\Delta \mathfrak{t}
\label{eq:22}\\
&=& +\left(\int_{t_1}^{t_2} F \frac{dq_{i+1} }{dt} dt  \right).
\end{eqnarray}

Thus $\delta w$ \textit{per unit time} is given by 
\begin{equation}
\left(\int_{t_1}^{t_2} F \frac{dq_{i+1} }{dt} dt\right)/n\Delta t
\label{eq:24}
\end{equation}
which is identical to $\delta w_{1\rightarrow 2}$ in Table (\ref{tab:1}) and
(\ref{eq:7b}). We note therefore complete concordance  between the two models.
We have modeled transitions here in terms of one particle that exhibits
recoverable transitions. Relative to the work transitions, one might wish to
also characterize the lower exchange temperature and  heat energy $Q_a$ that
allows for work to be performed on the adjacent elements. 
Over $n$  time intervals $dt$, we have 
 \begin{equation}
Q_a\left|_n\right.= \frac{\int_{t_1}^{t_2} \left( k.e.(i) -F \frac{dq_{i+1}}{dt}
dt\Delta \mathfrak{t} \right)}{dt}.
\label{eq:25}
\end{equation}
 Then $<Q_a>$ for one  interval is 
\begin{equation} \label{eq:26}
	<Q_a>=\frac{\int k.e.(i)}{ndt} - \left( \frac{\int F 
\frac{dq_{i+1}}{dt} dt}{n}\right).
\end{equation}
Hence, 
\begin{eqnarray}
<Q_a> &=& \left\{<Q_b>- \delta w_{1\rightarrow 2} \times dt\right\} = \alpha\\
&\mbox{and}&
<T_a>=2\alpha .
\label{eq:27}
\end{eqnarray}
The results for $<T_a>$ are provided in  Table \ref{tab:1}. We observe that it
is lower, as it should be, and this temperature is associated with the particle
$i$. Thus in any one location, we observe that we can evoke 2 temperatures that
are consistent with the energy  transfer across the crystal and the extended
Zeroth law. 

\section{Results and Discussions}\label{sec:5}
\begin{figure}[htbp]\label{fig:1}
	\centering
width=6cm]{tp1.eps}
\includegraphics[width=10cm]{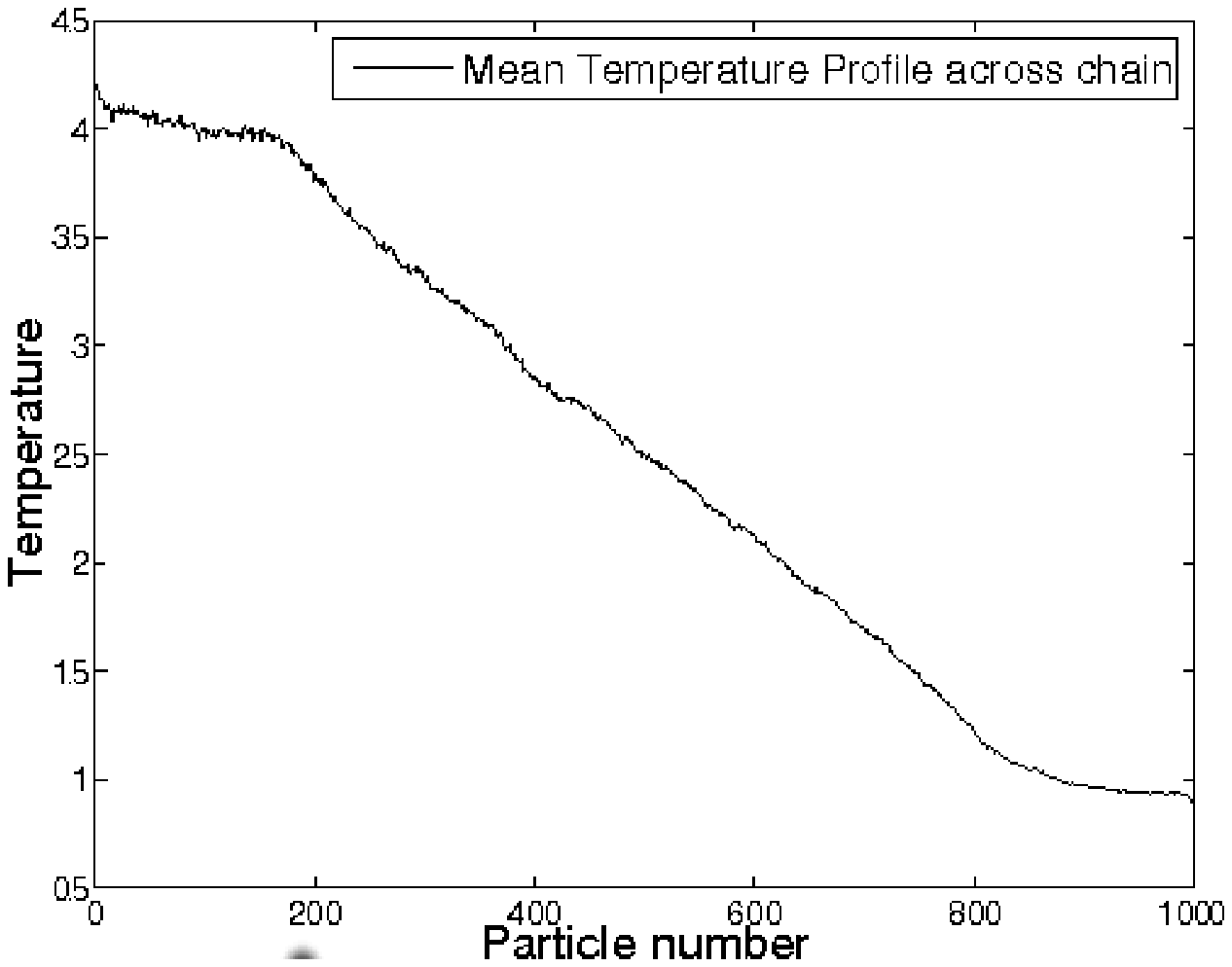} 
  \caption{Temperature Profile across chain. From Table \ref{tab:1}, the
\textit{u.s.d.} for the freely vibrating particles is of the order of .0.14
whereas the  thermalized particles  1-200 are maintained at the average
temperature T=0.40014E+01,  \usd  0.8E-02, and the colder themalized particles
800-1000  have  the average temperature T= 0.99987E+00, \usd   0.2E-02}
\end{figure}
The use of an anharmonic potential allows for the system to exhibit a near
linear temperature gradient as shown in Figure.\ref{fig:1}. The details of the
simulations are as follows:\newline
 all fluctuations  in quantities are expressed as the uncorrected standard
deviation \textit{u.s.d.} and the error $\pm$ are expressed in terms of \usd.
The E notation represents exponents to base 10. After  many successive
equilibration   runs amounting to about 500M timesteps (  where $dt=0.001$ for
the timestep  in reduced units), the production runs were initiated. The
integrations utilized  the well-established symplectic  Velocity Verlet  
algorithm of Swope, Anderson,Berens and Wilson \cite[p.81,eqs.3.17-3.21 of Allen
et al. reference]{allen2, swope1}, which is essentially second order. The
production runs  are for 100M timesteps, where a continuous sampling of 3M steps
 (constituting a dump)  are made over 20 dumps. The various statistics are
obtained over these 20 dump  values and the fluctuations expressed as   the
uncorrected standard  deviation (\usd ).  A current vogue in these studies is
the fifth-order  Runge-Kutta integrator algorithm \cite{tejal1}, where it is
arguable whether more accurate 
results necessarily obtain due to machine error accumulation; here we  wish to 
establish some principles that require sampling  a relatively larger portion of
phase space  that would be precluded by computational costs of more intensive 
computational algorithms that are only relevant for accurate determination over
a smaller region of phase space.

\begin{table}[ht] 
\centering
\begin{tabular} {c c c c c c         } \hline
       part.\# &  $\delta w_{2 \rightarrow 1} $ & $\delta w_{1\rightarrow 2}$&
temp. \#  & temp. \#+1 &$<T_a>$ \\ \hline
       
       250  & -0.21070E+00 &   0.21110E+00  &  0.35166E+01  &  0.35038E+01&
0.35162E+01  \\
       300  & -0.21031E+00  &  0.21046E+00  &  0.33159E+01  &  0.33037E+01 & 
0.33154E+01\\
       350  & -0.20958E+00  &  0.20976E+00  &  0.31287E+01   & 0.31141E+01&
0.31283E+01 \\ 
       
       450  & -0.20978E+00 &   0.20988E+00  &  0.27131E+01  &  0.26945E+01&
0.27127E+01 \\ 
       500 &  -0.21036E+00 &   0.21034E+00  &  0.24834E+01   & 0.24830E+01 &
0.24830E+01\\ 
       550  & -0.21185E+00  &  0.21182E+00  &  0.23192E+01  &  0.23078E+01 &
0.23187E+01 \\
        
       650  & -0.21076E+00  &  0.21086E+00  &  0.18816E+01  &  0.18692E+01
&0.18812E+01  \\
       700  & -0.21149E+00  &  0.21146E+00  &  0.16796E+01  &  0.16676E+01
&0.16792E+01 \\ 
       750  & -0.21236E+00  &  0.21231E+00  &  0.14668E+01   &
0.14619E+01&0.14664E+01  \\ \hline
\end{tabular}
\caption{The particle index number  \# is provided by the header on the
\textit{l.h.s.}, with all the other  variables (such as the $\delta w$'s)  fully
described in the text. The temperature of the particle \# and  that of the
adjacent one  on its  \textit{r.h.s.} with index \#+1 appears in columns 4 \& 5.
Column 6 are the results for $<T_a>$ given by (\ref{eq:27}).  The 
\textit{u.s.d} of the results for  the $\delta w$'s is approximately $0.12\times
10^{-1}$ and that of the temperatures $0.14\times 10^{0}$.  }
\label{tab:1}
\end{table}

The rate of  energy transfer into the 200 hot thermostated atoms at the left
hand side of the system is  21747E+00 $\pm $ 0.34577E-01 and for the $200$
colder atoms is    -0.21229E+00 $\pm $ 0.12843E-01. We find that the energy
transferred to the adjacent atoms in Table \ref{tab:1}, $\delta w_{1 \rightarrow
2}$  based on recoverability theory is in excellent quantitative agreement to
the independently determined energy flow into the thermostats in every instance.
We note that the reverse work of particle $i+1  \rightarrow i$ is negative
meaning that there is a gain of energy due to its own force field. 
The important point therefore is that one is  observing a type of one way
scattering of energy, from  $i$ to $i+1$ due to the  conservation of energy
because atom $i-1$ would have to scatter  energy into $i$ to compensate for the
loss of energy through the force field on the work done to $i+1$. 

\section{Conclusions}
Over the last two centuries, the main developments  in thermodynamics include
(a) the partitioning of the energy  into two distinct forms, work $W$ and heat
$Q$, from which the entropy is derived from the latter where the entropy
differential $dS$ for a closed system is $dS=dQ/T$, and (b) where the field of
statistical theory in concerned, these concepts were cast in terms of the
Liouville equation  and the associated Hamiltonian for both classical and
quantum systems. We have shown  the mathematical contradictions of both the
Liouville equation and  the time reversibility assumptions, both of which were
incorporated into the more modern versions of thermodynamical theories. Here we
have outlined a method of conflating both these quantities via the concept of
recoverable transitions. We have demonstrated its feasibility  for the case of
single particle scattering. Of great  interest, and of great challenge, is to
extend these models to  more than one particle, so that the kinetics and
thermodynamical profile 
of the system might be computed from these elementary 
considerations.  We also contradict the view that the Fourier law is a locally
defined process, unrelated to the Second law or which cannot be deduced from the
Second law, a claim central to the work of Benofy and Quay  in their theory of
thermoelectric and thermomagnetic effects\cite{bq}, by framing elementary
theoretical  propositions that are then tested out numerically in simulations.
Benofy and Quay interpreted the  Second law as  pertaining  to globally coupled
heat-work transitions, whereas (Fourier) heat conduction was viewed as local and
also as only one component of the energy definition, and therefore had only
local significance.  
It seems  feasible  from the preliminary results obtained here that
thermodynamical processes, both equilibrium  and otherwise may be expressed in
terms of a single concept where equilibrium and non-equilibrium processes might 
appear as limiting cases of this singular interpretation. What is of interest in
these initial forays is the nature of the representation; here we are able to 
model systems that seem to conform to recoverable trajectories for single
particle interactions. Questions that immediately suggest themselves are 
whether these models are unique, and if not, do they imply  a multiplicity of
modes  that recoverability theory can accommodate to interpret thermodynamical
phenomena relative to the unified concept being proposed.
 \begin{theacknowledgments}
  This work was supported by University of Malaya Grant UMRG(RG293/14AFR). I
thank the organizers, in particular Seenith Sivasundaram for notifying me of
this excellent conference series and Eva Klasik for her remarkable efficiency
and goodwill  in dealing with conference issues.
\end{theacknowledgments}
\bibliographystyle{aipproc} 
\bibliography{narvikbib}

\begin{thebibliography}{23}
\expandafter\ifx\csname natexlab\endcsname\relax\def\natexlab#1{#1}\fi
\providecommand{\enquote}[1]{``#1''}
\expandafter\ifx\csname url\endcsname\relax
  \def\url#1{\texttt{#1}}\fi
\expandafter\ifx\csname urlprefix\endcsname\relax\def\urlprefix{URL }\fi
\providecommand{\eprint}[2][]{\url{#2}}

\bibitem[Jesudason(1991)]{cgj14}
C.~G. Jesudason, \emph{Indian J. Pure Ap. Phy.} \textbf{29}, 163--182 (1991).

\bibitem[Jesudason.(2011)]{cgj16}
C.~G. Jesudason., \enquote{I. time reversibility concepts, the second law and
  irreversible thermodynamics,} in \emph{SECOND LAW OF THERMODYNAMICS: STATUS
  AND CHALLENGES}, edited by D.~P. Sheehan, A.A.A.S., American Institute of
  Physics, 2011, vol. 1411, pp. 292--307.

\bibitem[Jesudason(1999{\natexlab{a}})]{cgj4}
C.~Jesudason, \emph{Apeiron} \textbf{6(1-2)}, 9--24 (1999{\natexlab{a}}),
  url:www.redshift.vif.com.

\bibitem[Jesudason(1999{\natexlab{b}})]{cgj5}
C.~Jesudason, \emph{Apeiron} \textbf{6(1-2)}, 172--185 (1999{\natexlab{b}}),
  url:www.redshift.vif.com.

\bibitem[Landsberg(1990)]{lands1}
P.~T. Landsberg, \emph{Thermodynamics and Statistical Mechanics}, Dover Publ.
  Inc., New York, 1990.

\bibitem[Carnot(1897)]{carnot1}
N.~L.~S. Carnot, \emph{Reflections on the motive power of heat}, John Wiley \&
  Sons, Chapman \& Hall, New York and London, 1897, second revised edn.,
  translation of 1824 memoir of Carnot from the French, edited by R. H.
  Thurston.

\bibitem[Zemansky and Dittman(1997)]{zemansky1}
M.~W. Zemansky, and R.~H. Dittman, \emph{HEAT AND THERMODYNAMICS}, McGraw-Hill
  International, New York, 1997, 7 edn.

\bibitem[Carath\'{e}odory(1909)]{cara1}
C.~Carath\'{e}odory, \emph{Math. Ann.} \textbf{67}, 355--386 (1909), a special
  translation of this work into English was referred to in "The Second Law of
  Thermodynamics" (Dowden, Hutchinson and Ross, Stroudsburg, Penn., 1976)
  edited by J. Kestin, Chapter 12 .

\bibitem[Clapeyron(1976)]{clap1}
E.~Clapeyron, \enquote{Memoir on the motive power of HEAT,} in \emph{The Second
  Law of Thermodynamics}, edited by J.~Kestin, Dowden, Hutchinson \& Ross,
  Inc., Stroudsburg,Penn., 1976, vol.~5, pp. 36--51.

\bibitem[Jesudason(2003)]{cgj12}
C.~G. Jesudason, \emph{Stoch Anal Appl} \textbf{21(5)}, 1097--1114 (2003),
  erratum with publisher's apology for wrong typesetting of key equations in
  22(4), pp.1131-1132 (2004).

\bibitem[de~Groot and Mazur(1984)]{degroot1}
S.~de~Groot, and P.~Mazur, \emph{Non-equilibrium thermodynamics}, Dover Publ.
  Inc., New York, 1984, first edn.

\bibitem[Carath\'{e}odory(1976)]{cara2}
C.~Carath\'{e}odory, \enquote{Investigations into the foundations of
  thermodynamicS,} in \emph{The Second Law of Thermodynamics}, edited by
  J.~Kestin, Dowden, Hutchinson \& Ross, Inc., Stroudsburg, Penn., 1976, vol.~5
  of \emph{Benchmark Papers on Energy}, chap.~12, pp. 229--256.

\bibitem[Hafskjold and Ikeshoji(1995)]{Haf7}
B.~Hafskjold, and T.~Ikeshoji, \emph{Fluid Phase Equilibria} \textbf{104},
  173--184 (1995).

\bibitem[Rieder et~al.(1967)]{rieder1}
Z.~Rieder, J.~L. Lebowitz, and E.~Lieb, \emph{J. Math. Phys} \textbf{8},
  1073--1078 (1967).

\bibitem[Bonetto et~al.(2000)]{lebo2}
F.~Bonetto, J.~Lebowitz, and L.~Rey-Bellet, \emph{arXiv:math-ph/0002052v1}
  (2000).

\bibitem[Shah and Gajjar(2013)]{tejal1}
T.~N. Shah, and P.~N. Gajjar, \emph{Commun. Theor. Phys.} \textbf{59}, 361--364
  (2013).

\bibitem[Press et~al.(2007)]{nrc}
W.~Press, S.~Teukolsky, W.~Vetterling, and B.~Flannery, \emph{Numerical Recipes
  -The Art of Scientific Computing\hspace{.2cm}}, Cambridge University Press,
  2007, third edn.

\bibitem[Yakowitz and Szidarovsky(1990)]{yak1}
S.~Yakowitz, and F.~Szidarovsky, \emph{An Introduction to Numerical
  Computations\hspace{.2cm}}, Maxwell Macmillan, New York, 1990.

\bibitem[Prosen and Robnik(1992)]{prosen1}
T.~Prosen, and M.~Robnik, \emph{J. Phys. A: Math. Gen.} \textbf{25}, 3449--3472
  (1992).

\bibitem[DePree and Swartz(1988)]{depree1}
J.~DePree, and C.~Swartz, \emph{Introduction to Real Analysis}, John Wiley \&
  Sons, New York, 1988.

\bibitem[Allen and Tildesley(1990)]{allen2}
M.~P. Allen, and D.~J. Tildesley, \emph{Computer Simulation of Liquids},
  Clarendon Press, Oxford, 1990, first edn.

\bibitem[Swope et~al.(1982)]{swope1}
W.~Swope, H.~C. Anderson, P.~H. Berens, and K.~R. Wilson, \emph{J. Chem. Phys.}
  \textbf{76}, 637--49 (1982).

\bibitem[Benofy and \mbox{Quay, S.J.}(1982)]{bq}
S.~J. Benofy, and P.~M. \mbox{Quay, S.J.}, \enquote{Fourier inequality as a
  principle of thermodynamics, with applications to magnetothermoelectric
  effects,} in \emph{Physics as natural philosophy : essays in honor of Laszlo
  Tisza on his seventy-fifth birthday}, edited by L.~Tisza, A.~Shimony, and
  H.~Feshbach, M.I.T. Press, 1982, chap.~1, pp. 7--24.

\end{thebibliography}
\end{document}